\documentstyle[12pt]{article}
\begin{document}
\title{On the Weak-Field Approximation in Generalized Scalar-Tensor Gravities}
\author{M. E. X. Guimar\~aes$^1$, L. P. Colatto$^2$ and F. B. Tourinho$^2$ \\
\mbox{\small{1. Universidade de Bras\'{\i}lia, Depto. de Matem\'atica, 
CEP: 70910-900, Bras\'{\i}lia - DF, Brazil}} \\
\mbox{\small{2. Universidade de Bras\'{\i}lia, Instituto de F\'{\i}sica,
CEP: 70910-900 Bras\'{\i}lia - DF, Brazil}} \\
\mbox{\small{ E-mail: emilia@mat.unb.br, \,\, colatto@fis.unb.br, \,\, felipetourinho@ambr.br}}}
\date{}
\maketitle
\begin{abstract}
In the paper \cite{bar}, Barros and Romero demonstrated that, in the weak-field approximation, solutions to the Brans-Dicke equations are related to the solutions of General Relativity for the same matter distributions. In the present work, we enphasize this result and we extend it to for generalized scalar-tensor theories in which the parameter $\omega$ is no longer a constant but an arbitrary 
function of the (gravitational) scalar field. 
\end{abstract}

\section{Introduction:}

In the paper \cite{bar}, Barros and Romero developed a method which allowed one to obtain solutions in the Brans-Dicke theory \cite{bra} from the corresponding solutions in 
General Relativity theory for the same matter content when both theories are considered in the weak-field approximation. The aim of this work is to show that this is a global feature of a class of more general scalar-tensor theories of gravity in which the parameter $\omega$ is an arbitrary function of the scalar field $\tilde{\Phi}$ \cite{wa}. 
Since, at present, the only known theory which treats gravity consistently 
with quantum mechanics is the string theory \cite{gre} and since all versions of this theory naturally predicts a scalar partner for the pure tensor metric, it seems worthwhile to analyse the behaviour of matter contents in the context of a scalar-tensor gravity. Hence, our main contribution is to provide a way to derive solutions in scalar-tensor theories from their General Relativity partners straightforwardly, at least 
in the weak-field approximation. 

This work is presented as follows. In the section 2 we present our model and its field equations and extend the method introduced in the ref. \cite{bar}. In section 3, we illustrate our result with some examples of matter distributions such as topological defects. In section 4, we end with some conclusions. 

\section{The Linearised Field Equations in Scalar-Tensor Gravities:}

We start with the action in the Jordan-Fierz frame:
\begin{equation}
{\cal S}= \frac{1}{16\pi} \int d^4 x \sqrt{-\tilde{g}} \left[ \tilde{R}\tilde{\Phi} - \frac{\omega(\tilde{\Phi})}{\tilde{\Phi}}\partial^{\mu}\tilde{\Phi}\partial_{\mu}\tilde{\Phi}\right] + {\cal S}_{m}[\Psi_m , \tilde{g}_{\mu\nu}] ,
\end{equation}
$\tilde{g}_{\mu\nu}$ is the physical metric which contains both scalar and tensor degrees of freedom, $\tilde{R}$ is the curvature scalar associated to it and ${\cal S}_{m}$ is the action for general matter fields which, by now, is left arbitrary.

By varying action (1) with respect to the metric $\tilde{g}_{\mu\nu}$ and to the scalar field $\tilde{\Phi}$ we obtain the ``modified" Einstein equations and a wave equation for $\tilde{\Phi}$:
\begin{eqnarray}
\tilde{R}_{\mu\nu} - \frac{1}{2}\tilde{g}_{\mu\nu}\tilde{R} & = & \frac{8\pi}{\tilde{\Phi}}\tilde{T}_{\mu\nu} + \frac{\omega(\tilde{\Phi})}{\tilde{\Phi}}\left[\partial_{\mu}\tilde{\Phi}
\partial_{\nu}\tilde{\Phi} -\frac{1}{2}\tilde{g}_{\mu\nu}\partial^{\alpha}
\tilde{\Phi}\partial_{\alpha}\tilde{\Phi} \right] \nonumber \\
& & +  \frac{1}{\tilde{\Phi}}\left(\nabla_{\nu}\tilde{\Phi}_{,\mu} - 
\tilde{g}_{\mu\nu}\Box_{\tilde{g}}\tilde{\Phi} \right) , \nonumber \\
\Box_{\tilde{g}}\tilde{\Phi} & = & \frac{1}{2\omega(\tilde{\Phi}) + 3} 
\left[ 8\pi \tilde{T} - \frac{d\omega}{d\tilde{\Phi}}\partial_{\mu}\tilde{\Phi}\partial^{\mu}\tilde{\Phi} \right] \\
\nabla_{\mu}\tilde{T}^{\mu}_{\nu} & = & 0 
\end{eqnarray}
where
\[
\tilde{T}_{\mu\nu} = \frac{2}{\sqrt{-\tilde{g}}} \frac{\delta {\cal S}_m}{\delta \tilde{g}^{\mu\nu}}
\]
is the energy-momentum tensor of the matter content and $\tilde{T} \equiv \tilde{T}^{\mu}_{\mu}$ is its trace. Clearly, if $\tilde{T}$ vanishes and $\tilde{\Phi}$ is a constant, equations (2) reduce to the usual Einstein's equations if we identify $G$ with the inverse of the scalar field, e.g., 
$G = 1/\tilde{\Phi}$. Hence, any exact solution of Einstein's equations with a trace-free matter source will also be a particular exact solution of the scalar-field with $\tilde{\Phi}$ constant. Of course, this particular solution will not be the general solution for the matter content \cite{baw}. 
 
Before proceeding with the linearisation, let us re-write action (1) in terms of the Einstein (conformal) frame in which the kinematic terms of tensor and scalar do not mix:
\begin{equation}
{\cal S} = \frac{1}{16\pi G^*} \int \sqrt{g} \left[ R - 2g^{\mu\nu} \partial_{\mu}\phi \partial_{\nu}\phi \right] + {\cal S}_m [\Psi_m , A^2(\phi)g_{\mu\nu}] ,
\end{equation}
where $g_{\mu\nu}$ is a pure rank-2 metric tensor and $R$ is the curvature scalar associated to it. 

Action (4) is obtained from (1) by a conformal transformation 
\begin{equation}
\tilde{g}_{\mu\nu} = A^2(\phi) g_{\mu\nu} , 
\end{equation}
and by a redefinition of the quantities
\begin{equation}
G^* A^2(\phi) = \frac{1}{\tilde{\Phi}}
\end{equation}
$G^*$ is a ``bare" gravitational constant, and
\begin{equation}
\alpha(\phi) = \frac{ \partial{\ln A(\phi)}}{\partial \phi} = \frac{1}{(2\omega(\tilde{\Phi}) + 3)^{1/2}}
\end{equation}
which can be interpreted as the (field-dependent) coupling strength between matter and scalar field. 

In the conformal frame, eqs. (2) are written in a more convenient form:
\begin{eqnarray}
R_{\mu\nu} - \frac{1}{2}g_{\mu\nu}R & = & 8\pi G^* T_{\mu\nu} + 2\partial_{\mu}\phi\partial_{\nu}\phi - g_{\mu\nu}g^{\alpha\beta}\partial_{\alpha}\phi\partial_{\beta}\phi \nonumber \\
\Box_{g}\phi & = & - 4\pi G^* \alpha(\phi) T .
\end{eqnarray}

From eq. (5), it is clear that we can relate quantities from both frames such that $\tilde{T}^{\mu\nu} = A^{-6}(\phi) T^{\mu\nu}$ and $\tilde{T}^{\mu}_{\nu} = A^{-4}(\phi) T^{\mu}_{\nu}$. Let us expand the fields to first order in the parameter $G_0 = G^*A^2(\phi_0)$, we then obtain
\begin{eqnarray}
\label{ee}
g_{\mu\nu} & = & \eta_{\mu\nu} + h_{\mu\nu} \nonumber \\
\phi & = & \phi_0 + \phi_{(1)} \\
A(\phi) & = & A(\phi_0)[ 1 + \alpha(\phi_0)\phi_{(1)}] \nonumber \\
T^{\mu}_{\nu} & = & T^{\mu}_{(0)\nu} + T^{\mu}_{(1)\nu} \nonumber 
\end{eqnarray}
Therefore, eqs. (8) reduce to:
\begin{eqnarray}
\label{li}
\nabla^2 h_{\mu\nu} & = & 16\pi G^* (T_{(0)\mu\nu} - 
\frac{1}{2} \eta_{\mu\nu}T_{(0)}) \\
\nabla^2 \phi_{(1)} & = & 4\pi G^* \alpha(\phi_0)T_{(0)} \nonumber 
\end{eqnarray}

In this approximation, $T^{(0)}_{\mu\nu}$ is the energy-momentum tensor at zeroth-order in the conformal frame. Its relation to the (physical) energy-momentum tensor at zeroth-order in the Jordan-Fierz frame is given by 
$T^{(0)}_{\mu\nu}=A^2(\phi_0)\tilde{T}^{(0)}_{\mu\nu}$. In this way, the first equation in the system (10) is the Einstein's equation in the weak-field approximation regime. 

Now, $\tilde{g}_{\mu\nu} = A^2(\phi)[\eta_{\mu\nu} + h_{\mu\nu}]$. Therefore, using the approximation (\ref{ee}), we have:
\begin{equation}
\tilde{g}_{\mu\nu} = A^2(\phi_0)[ 1 + 2\alpha(\phi_0)\phi_{(1)}][\eta_{\mu\nu} + h_{\mu\nu}]
\end{equation}
We can see then that the problem of finding the metric in the scalar-tensor gravities may reduce to find the metric in Einstein's gravity for the same matter distribution. In the next section, we will illustrate our proposal with some examples, such as topological defects. 

\section{First Order Solutions:}

In this section, we will apply the method developed in the previous section to the cases 
of a cosmic string, a domain wall and a monopole, respectively.

\subsection{The Domain Wall Solution:}

In what follows, we will consider a static domain wall with neglegible width lying in the 
$yz$-plane in the weak-field approximation. Therefore, 
\begin{equation}
\label{dw}
T^{\mu}_{(0)\nu} = A^4(\phi_0)\sigma \delta(x) diag (1,0,1,1)
\end{equation}
in the cartesian coordinate system $(t,x,y,z)$.  $\sigma$ is the wall's 
surface energy density. In our convention, the metric signature is $-2$.  

Let us begin by solving the equation for the dilaton field $\phi_{(1)}$ 
in (\ref{li}):
\begin{eqnarray}
\nabla^2\phi_{(1)} & = & 12\pi \sigma G_0 A^2(\phi_0)\alpha(\phi_0)\delta(x) \nonumber \\
\phi_{(1)} & = & 6 \pi \sigma G_0\alpha(\phi_0)\mid x\mid ,
\end{eqnarray}
where $G_0 \equiv G^* A^2(\phi_0)$. 

Now, the linearised Einstein's equation in (\ref{li}) with source given by 
(\ref{dw}) are just the same as in Vilenkin's paper\cite{vil2}, except that 
in our case the metric is multiplied by the factor $A^2(\phi)$ linearised. 
Therefore, we have (to first order in $G_0$):
\[
ds^2 = A^2(\phi_0) \left[ 1 + 4\pi \sigma G_0  |x|(3 \alpha^2 (\phi_0) - 1) \right] [dt^2 - dx^2 - dy^2 - 
dz^2] .\]
The factor $A^2(\phi_0)$ appearing in the above expression can be absorbed by 
a redefinition of the coordinates $(t,x,y,z)$. We finally, then, obtain \cite{be}:
\begin{equation}
\label{me}
ds^2 =  \left[ 1 + 4\pi \sigma G_0  |x|  (3 \alpha^2 (\phi_0) - 1) \right] 
[dt^2 - dx^2 - dy^2 - dz^2] .
\end{equation}
This is the line element corresponding to a domain wall in the framework
of scalar-tensor gravity in the weak-field approximation. It is very illustrative to  consider a particular form for the arbitrary function $A(\phi)$, corresponding to the Brans-Dicke theory, $A(\phi) = e^{\alpha \phi}$ , with $\alpha^{2} = \frac{1}{2 \omega + 3}, (\omega = cte)$. In this case, we have that  $ G^* A^{2}(\phi_{0})= G_{0}= \left( \frac{2 \omega + 3}{2 \omega + 4} \right) G_{eff}$ \cite{bra}  where $G_{eff}$ is the Newtonian constant. Therefore, metric (\ref{me}) reduces to the same as in Barros and Romero \cite{bar,br}. 

\subsection{The Cosmic String Solution:}

Let us consider a static string lying in the $z$-axis. 
In this case, the energy-momentum at zeroth-order is given by:
\begin{equation}
\label{string}
T^{\mu}_{(0)\nu} = A^4(\phi_0)\mu \delta(\rho) diag(1,0,0,1)
\end{equation}
in a cylindrical coordinate system $(t,r,\theta,z)$. $\mu$ is the string's linear energy density. 

Again, let us begin by solving the equation for the dilaton field $\phi_{(1)}$ in (\ref{li}):
\begin{equation}
\phi_{(1)} = 4 G_0 A^2(\phi_0) \mu \alpha(\phi_0) \ln \rho
\end{equation}

The procedure here to compute the metric is the same as exposed in the case of a domain wall, since we have to solve Einstein's equations at linear order. Therefore, we have \cite{mexg}:
\begin{equation}
ds^2 = \left[1 + 8 G_0 \mu \alpha^2(\phi_0) \ln \rho\right]\left[ dt^2 - dz^2- d\rho^2 - (1- 8 G_0 \mu)  \rho^{2}d\theta^2 \right] 
\end{equation}
Again, this result reduces to the metric found by Barros and Romero \cite{bar,br} in the particular case of Brans-Dicke theory.

\subsection{The Monopole Solution:}

Let us consider now a global monopole with total mass $m = 4\pi \eta^2 R$, where $\eta$ is the energy scale of the symmetry breaking and $R$ is the cut-off radius. Then, the energy-momentum tensor in spherical coordinates $(t,r, \theta,\varphi)$ is given by:
\begin{equation}
T^{\mu}_{(0)\nu} = A^4(\phi_0)\frac{\eta^2}{r^2}diag(1,1,0,0)
\end{equation}

Solving the equation for the dilaton field, we have
\begin{equation}
\phi_{(1)} = 8\pi G_0 A^2(\phi_0)\alpha(\phi_0)\eta^2\ln r
\end{equation}
In this way, we can easily obtain the metric for a global monopole in the weak-field approximation from the solution of Vilenkin and Barriola \cite{vil1}:
\begin{equation}
ds^2 = [ 1 + 16\pi G_0 A^2(\phi_0)\alpha^2(\phi_0)\eta^2\ln r] [dt^2 - dr^2 - (1- 8\pi G_0\eta^2) (d\theta^2 + \sin^2\theta d\varphi^2)]
\end{equation}
Again, we reduce to the same result found previously by Barros and Romero \cite{br} in the case of Brans-Dicke.

\section{Conclusions:}

Our main result is to show that there is a correspondence between the metric solution in scalar-tensor gravities and the metric solution in Einstein's gravity for the same matter distribution, in the weak-field approximation. 
Indeed, we have demonstrated that the linearized metric in Einstein's gravity is multiplied by a conformal factor which depends on the solution of the dilaton's equation for each matter content.   

To conclude, we briefly mention that the gravitational field generated by 
topological defects in scalar-tensor gravities present many intersting features, already at linear order. First of all, light propagates in the same way as in the General Relativity case. However, defects in the scalar-tensor gravities exert a gravitational force on massive test particles and this fact leads to interesting consequences, such as for instance a perturbation in the particles' velocity implying in the formation of wakes \cite{ma} by moving strings and the generation of a current inside strings \cite{pet}, a new feature that still deserves further analysis \cite{tou}. 

\section*{Acknowledgements:}

L. P. C. thanks CAPES for a financial support. F. B. Tourinho thanks CNPq for a grant in the context of the program PIBIC/UnB.


\begin{thebibliography}{99}
\bibitem{bar}A. Barros and C. Romero, {\em Phys. Lett. A}{\bf 245} (1998) 31. 
\bibitem{bra}C. Brans and R. H. Dicke, {\em Phys. Rev.}{\bf 124} (1961) 925.
\bibitem{wa}P. G. Bergmann, {\em Int. J. Theor. Phys.}{\bf 1} (1969) 25; R. V. Wagoner, {\em Phys. Rev. D}{\bf 1} (1970) 3209; K. Nordtverdt Jr., {\em Ap. J.}{\bf 161} (1970) 1059.
\bibitem{gre}M. B. Green, J. H. Schwarz and E. Witten, {\em Superstring Theory} (Cambridge University Press, 1987).
\bibitem{baw}J. D. Barrow, {\em Phys. Rev. D}{\bf 47} (1993) 5329.
\bibitem{vil2}A. Vilenkin, {\em Phys. Rev. D}{\bf 23} (1981) 852.
\bibitem{br}A. Barros  and C. Romero {\it J. Math. Phys.} {\bf 36} (1995) 5800. 
\bibitem{be}V. B. Bezerra, L. P. Colatto, M. E. X. Guimar\~aes and R. T. Muniz Filho, {\em gr-qc/0104038, submitted for publication} (2001).  
\bibitem{mexg}M. E. X. Guimar\~aes, {\em Class. Quantum Grav.}{\bf 14} 
(1997) 435.
\bibitem{vil1}M. Barriola and A. Vilenkin, {\em Phys. Rev. Lett.}{\bf 63} (1989) 341.
\bibitem{ma}S. R. M. Masalskiene and M. E. X. Guimar\~aes, {\em Class. 
Quantum Grav.}{\bf 17} (2000) 3055. 
\bibitem{pet}V. C. de Andrade, P. Peter and M. E. X. Guimar\~aes, {\em gr-qc/0101025, submitted for publication} (2001). 
\bibitem{tou}M. E. X. Guimar\~aes, P. Peter and F. B. Tourinho, {\em in preparation} (2001). 

\end{thebibliography}
\end{document}